\documentclass[12pt,preprint]{aastex}
\usepackage{lscape}
\usepackage{rotating}

\def\kms{{\rm km}\,{\rm s}^{-1}}

\def\max{{\rm max}}

\def\e{{\rm E}}

\begin{document}
\title{MOA-2010-BLG-523: ``Failed Planet'' = 
RS CVn Star\footnote{Based on observations made with the 
European Southern Observatory telescopes, Program ID 85.B-0399(I)}}
\author{
A.~Gould\altaffilmark{1},
J.C.~Yee\altaffilmark{1},
I.A.~Bond\altaffilmark{2},
A.~Udalski\altaffilmark{3},
C.~Han\altaffilmark{4},
U.G.~J{\o}rgensen\altaffilmark{5,6},
J.~Greenhill\altaffilmark{7},
Y.~Tsapras\altaffilmark{8,9},
M.H.~Pinsonneault\altaffilmark{1},
T.~Bensby\altaffilmark{10}
\and
W.~Allen\altaffilmark{11},
L.A.~Almeida\altaffilmark{12},
M.~Bos\altaffilmark{13},
G.W.~Christie\altaffilmark{14},
D.L.~DePoy\altaffilmark{15},
Subo~Dong\altaffilmark{16,1},
B.S.~Gaudi\altaffilmark{1},
L.-W.~Hung\altaffilmark{1},
F.~Jablonski\altaffilmark{12},
C.-U.~Lee\altaffilmark{17},
J.~McCormick\altaffilmark{18},
D.~Moorhouse\altaffilmark{19},
J.A.~Mu\~{n}oz\altaffilmark{20},
T.~Natusch\altaffilmark{14,21},
M.~Nola\altaffilmark{19},
R.W.~Pogge\altaffilmark{1},
J.~Skowron\altaffilmark{1},
G.~Thornley\altaffilmark{19}\\
(The $\mu$FUN Collaboration),\\
F.~Abe\altaffilmark{22},
D.P.~Bennett\altaffilmark{23,24},
C.S.~Botzler\altaffilmark{25},
P.~Chote\altaffilmark{26},
M.~Freeman\altaffilmark{25},
A.~Fukui\altaffilmark{27},
K.~Furusawa\altaffilmark{22},
P.~Harris\altaffilmark{26},
Y.~Itow\altaffilmark{22},
C.H.~Ling\altaffilmark{2},
K.~Masuda\altaffilmark{22},
Y.~Matsubara\altaffilmark{22},
N.~Miyake\altaffilmark{22},
K.~Ohnishi\altaffilmark{28},
N.J.~Rattenbury\altaffilmark{25},
To.~Saito\altaffilmark{29},
D.J.~Sullivan\altaffilmark{26},
T.~Sumi\altaffilmark{30,22},
D.~Suzuki\altaffilmark{30},
W.L.~Sweatman\altaffilmark{2},
P.J.~Tristram\altaffilmark{31},
K. Wada\altaffilmark{30},
P.C.M.~Yock\altaffilmark{25}\\
(The MOA Collaboration),\\
M.K.~Szyma{\'n}ski\altaffilmark{3},
I.~Soszy{\'n}ski\altaffilmark{3},
M.~Kubiak\altaffilmark{3},  
R.~Poleski\altaffilmark{3},
K.~Ulaczyk\altaffilmark{3},
G.~Pietrzy{\'n}ski\altaffilmark{3,32}, 
{\L}.~Wyrzykowski\altaffilmark{3,33}\\
(The OGLE Collaboration),\\
K.A.~Alsubai\altaffilmark{34},
V.~Bozza\altaffilmark{35,36}, 
P.~Browne\altaffilmark{37,38}, 
M.J.~Burgdorf\altaffilmark{39,40},
S.~Calchi Novati\altaffilmark{35,41},
P.~Dodds\altaffilmark{37},
M.~Dominik\altaffilmark{37,42,38}, 
F.~Finet\altaffilmark{43},
T.~Gerner\altaffilmark{44},
S.~Hardis\altaffilmark{5},
K.~Harps{\o}e\altaffilmark{5,6}, 
F.V.~Hessman\altaffilmark{45},
T.C.~Hinse\altaffilmark{5,46,17},
M.~Hundertmark\altaffilmark{37,45},
N.~Kains\altaffilmark{47,37,38}, 
E.~Kerins\altaffilmark{48},
C.~Liebig\altaffilmark{37,44}, 
L.~Mancini\altaffilmark{35,49},
M.~Mathiasen\altaffilmark{5}, 
M.T.~Penny\altaffilmark{1,48}, 
S.~Proft\altaffilmark{44}, 
S.~Rahvar\altaffilmark{50,51},
D.~Ricci\altaffilmark{43}, 
K.C.~Sahu\altaffilmark{52},
G.~Scarpetta\altaffilmark{35,36}, 
S.~Sch\"{a}fer\altaffilmark{45}, 
F.~Sch\"{o}nebeck\altaffilmark{44},
C.~Snodgrass\altaffilmark{53,54,38}, 
J.~Southworth\altaffilmark{55},
J.~Surdej\altaffilmark{45}, 
J.~Wambsganss\altaffilmark{44}\\ 
(The MiNDSTEp Consortium)\\
R.A.~Street\altaffilmark{8},
K.~Horne\altaffilmark{37},
D.M.~Bramich\altaffilmark{47},
I.A.~Steele\altaffilmark{56}\\
(The RoboNet Collaboration),\\
M.D.~Albrow\altaffilmark{57},
E.~Bachelet\altaffilmark{58},
V.~Batista\altaffilmark{1,59},
T.G.~Beatty\altaffilmark{1},
J.-P.~Beaulieu\altaffilmark{59},
C.S.~Bennett\altaffilmark{60},
R.~Bowens-Rubin\altaffilmark{61},
S.~Brillant\altaffilmark{62},
J.A.R.~Caldwell\altaffilmark{63},
A.~Cassan\altaffilmark{59},
A.A.~Cole\altaffilmark{7},
E.~Corrales\altaffilmark{59},
C.~Coutures\altaffilmark{59},
S.~Dieters\altaffilmark{7},
D.~Dominis Prester\altaffilmark{64},
J.~Donatowicz\altaffilmark{65},
P.~Fouqu\'e\altaffilmark{58},
C.B.~Henderson\altaffilmark{1},
D.~Kubas\altaffilmark{62,59},
J.-B.~Marquette\altaffilmark{59},
R.~Martin\altaffilmark{66},
J.W.~Menzies\altaffilmark{67},
B.~Shappee\altaffilmark{1},
A.~Williams\altaffilmark{66},
J.~van Saders\altaffilmark{1},
M.~Zub\altaffilmark{44},\\
(The PLANET Collaboration)
}

\altaffiltext{1}{Department of Astronomy, Ohio State University, 140 West 18th Avenue, Columbus, OH 43210, USA}
\altaffiltext{2}{Institute for Information and Mathematical Sciences, Massey University, Private Bag 102-904, Auckland 1330, New Zealand}
\altaffiltext{3}{Warsaw University Observatory, Al. Ujazdowskie 4, 00-478 Warszawa, Poland}
\altaffiltext{4}{Department of Physics, Chungbuk National University, Cheongju 361-763, Korea}
\altaffiltext{5}{Niels Bohr Institutet, K{\o}benhavns Universitet, Juliane Maries Vej 30, 2100 Copenhagen, Denmark}
\altaffiltext{6}{Centre for Star and Planet Formation, Geological Museum, {\O}ster Voldgade 5, 1350 Copenhagen, Denmark}
\altaffiltext{7}{University of Tasmania, School of Mathematics and Physics, Private Bag 37, Hobart, TAS 7001, Australia}
\altaffiltext{8}{Las Cumbres Observatory Global Telescope Network, 6740B Cortona Dr, Goleta, CA 93117, USA}
\altaffiltext{9}{School of Physics and Astronomy, Queen Mary University of London, Mile End Road, London, E1 4NS}
\altaffiltext{10}{Lund Observatory, Department of Astronomy and Theoretical physics, Box 43, SE-221\,00 Lund, Sweden}
\altaffiltext{11}{Vintage Lane Observatory, Blenheim, New Zealand}
\altaffiltext{12}{Divisao de Astrofisica, Instituto Nacional de Pesquisas Espaciais, Avenida dos Astronautas, 1758 Sao Jos\'e dos Campos, 12227-010 SP, Brasil}
\altaffiltext{13}{Molehill Astronomical Observatory, North Shore, New Zealand}
\altaffiltext{14}{Auckland Observatory, Auckland, New Zealand}
\altaffiltext{15}{Department of Physics and Astronomy, Texas A\&M University, College Station, Texas 77843-4242, USA}
\altaffiltext{16}{Institute for Advanced Study, Einstein Drive, Princeton, NJ 08540, USA}
\altaffiltext{17}{Korea Astronomy and Space Science Institute, 776 Daedukdae-ro, Yuseong-gu, Daejeon 305-348, Republic of Korea}
\altaffiltext{18}{Farm Cove Observatory, Centre for Backyard Astrophysics, Pakuranga, Auckland, New Zealand}
\altaffiltext{19}{Kumeu Observatory, Kumeu, New Zealand}
\altaffiltext{20}{Departamento de Astronomi{\'a} y Astrof{\'i}sica, Universidad de Valencia, E-46100 Burjassot, Valencia, Spain}
\altaffiltext{21}{AUT University, Auckland, New Zealand}
\altaffiltext{22}{Solar-Terrestrial Environment Laboratory, Nagoya University, Nagoya, 464-8601, Japan}
\altaffiltext{23}{Department of Physics, 225 Nieuwland Science Hall, University of Notre Dame, Notre Dame, IN 46556, USA}
\altaffiltext{24}{Also PLANET Collaboration}
\altaffiltext{25}{Department of Physics, University of Auckland, Private Bag 92-019, Auckland 1001, New Zealand}
\altaffiltext{26}{School of Chemical and Physical Sciences, Victoria University, Wellington, New Zealand}
\altaffiltext{27}{Okayama Astrophysical Observatory, National Astronomical Observatory, 3037-5 Honjo, Kamogata, Asakuchi, Okayama 719-0232, Japan}
\altaffiltext{28}{Nagano National College of Technology, Nagano 381-8550, Japan}
\altaffiltext{29}{Tokyo Metropolitan College of Aeronautics, Tokyo 116-8523, Japan}
\altaffiltext{30}{Department of Earth and Space Science, Graduate School of Science, Osaka University, 1-1 Machikaneyama-cho, Toyonaka, Osaka 560-0043, Japan}
\altaffiltext{31}{Mt. John University Observatory, P.O. Box 56, Lake Tekapo 8770, New Zealand}
\altaffiltext{32}{Universidad de Concepci\'on, Departamento de Astronom\'{\i}a, Casilla 160--C, Concepci\'on, Chile}
\altaffiltext{33}{Institute of Astronomy, University of Cambridge, Madingley Road, Cambridge CB3 0HA, UK}
\altaffiltext{34}{Qatar Foundation, P.O. Box 5825, Doha, Qatar}
\altaffiltext{35}{Dipartimento di Fisica "E.R.~ Caianiello", Universit\`{a} degli Studi di Salerno, Via Ponte Don Melillo, 84084 Fisciano, Italy}
\altaffiltext{36}{INFN, Sezione di Napoli, Italy}
\altaffiltext{37}{SUPA, University of St Andrews, School of Physics \& Astronomy, North Haugh, St Andrews, KY16 9SS, UK}
\altaffiltext{38}{Also RoboNet Collaboration}
\altaffiltext{39}{Deutsches SOFIA Institut but, HE Space Operations, Flughafenallee 26, 28199 Bremen, Germany}
\altaffiltext{40}{SOFIA Science Center, NASA Ames Research Center, Mail Stop N211-3, Moffett Field CA 94035, USA}
\altaffiltext{41}{Istituto Internazionale per gli Alti Studi Scientifici (IIASS), Vietri Sul Mare (SA), Italy}
\altaffiltext{42}{Royal Society University Research Fellow}
\altaffiltext{43}{Institut d'Astrophysique et de G\'{e}ophysique, All\'{e}e du 6 Ao\^{u}t 17, Sart Tilman, B\^{a}t.\ B5c, 4000 Li\`{e}ge, Belgium}
\altaffiltext{44}{Astronomisches Rechen-Institut, Zentrum f\"{u}r Astronomie der Universit\"{a}t Heidelberg (ZAH), M\"{o}nchhofstr.\ 12-14, 69120 Heidelberg, Germany}
\altaffiltext{45}{Institut f\"{u}r Astrophysik, Georg-August-Universit\"{a}t, Friedrich-Hund-Platz 1, 3707,7 G\"{o}ttingen, Germany}
\altaffiltext{46}{Armagh Observatory, College Hill, Armagh, BT61 9DG, Northern Ireland, UK}
\altaffiltext{47}{ESO Headquarters,Karl-Schwarzschild-Str. 2, 85748 Garching bei M\"{u}nchen, Germany}
\altaffiltext{48}{Jodrell Bank Centre for Astrophysics, University of Manchester, Oxford Road, Manchester, M13 9PL, UK}
\altaffiltext{49}{Max Planck Institute for Astronomy, K\"{o}nigstuhl 17, 619117 Heidelberg, Germany}
\altaffiltext{50}{Department of Physics, Sharif University of Technology, P.~O.\ Box 11155--9161, Tehran, Iran}
\altaffiltext{51}{Perimeter Institute for Theoretical Physics, 31 Caroline St. N., Waterloo ON, N2L 2Y5, Canada}
\altaffiltext{52}{Space Telescope Science Institute, 3700 San Martin Drive, Baltimore, MD 21218, USA}
\altaffiltext{53}{European Southern Observatory (ESO)}
\altaffiltext{54}{Max Planck Institute for Solar System Research, Max-Planck-Str. 2, 37191 Katlenburg-Lindau, Germany}
\altaffiltext{55}{Astrophysics Group, Keele University, Staffordshire, ST5 5BG, UK}
\altaffiltext{56}{Astrophysics Research Institute, Liverpool John Moores University, Liverpool CH41 1LD, UK}
\altaffiltext{57}{University of Canterbury, Department of Physics and Astronomy, Private Bag 4800, Christchurch 8020, New Zealand}
\altaffiltext{58}{IRAP, Universit\'e de Toulouse, CNRS, 14 Avenue Edouard Belin, 31400 Toulouse, France}
\altaffiltext{59}{UPMC-CNRS, UMR 7095, Institut d'Astrophysique de Paris, 98bis boulevard Arago, F-75014 Paris, France}
\altaffiltext{60}{Department of Physics, Massachussets Institute of Technology, 77 Mass. Ave., Cambridge, MA 02139, USA}
\altaffiltext{61}{Dept.\ of Earth, Atmospheric and Planetary Sciences, 54-1713, Massachusetts Institute of Technology, 77 Massachusetts Avenue, Cambridge, MA 02139, USA}
\altaffiltext{62}{European Southern Observatory, Casilla 19001, Vitacura 19, Santiago, Chile}
\altaffiltext{63}{McDonald Observatory, 16120 St Hwy Spur 78 \#2, Fort Davis, Texas 79734, USA}
\altaffiltext{64}{Department of Physics, University of Rijeka, Omladinska 14, 51000 Rijeka, Croatia}
\altaffiltext{65}{Technische Universit\"{a}t Wien, Wieder Hauptst. 8-10, A-1040 Vienna, Austria}
\altaffiltext{66}{Perth Observatory, Walnut Road, Bickley, Perth 6076, WA, Australia}
\altaffiltext{67}{South African Astronomical Observatory, P.O. Box 9 Observatory 7925, South Africa}

\begin{abstract}

The Galactic bulge source MOA-2010-BLG-523S 
exhibited short-term deviations from a standard microlensing
lightcurve near the peak of an $A_{\rm max}\sim 265$ high-magnification
microlensing event.  The deviations originally seemed consistent
with expectations for a planetary companion to the principal lens.
We combine long-term photometric monitoring with a previously
published high-resolution spectrum taken near peak to demonstrate
that this is an RS CVn variable, so that planetary microlensing
is not required to explain the lightcurve deviations.  This is
the first spectroscopically confirmed RS CVn star discovered in 
the Galactic bulge.

\end{abstract}


\section{{Introduction}
\label{sec:intro}}

High-magnification microlensing events provide a powerful tool for
planet detection, partly because planets are more likely to perturb
these events and partly because their high magnification (hence high
signal-to-noise ratio) allows even very small perturbations to be
detected.  However, non-microlensing flux variations are also
enhanced in these events.
In this paper we report on the discovery of an apparent planet
candidate that turned out instead to be a highly magnified active star
and discuss methods by which we identified and excluded this interloper.

Stars are intrinsically variable, and star spots can induce
substantial light curve variations in cool stars.  However, for most G and K
dwarfs this variability is manifested at a low level because magnetic
activity decays quickly with age.  Late M dwarfs can remain active
for a Hubble time, but they are faint and will not be common
microlensing sources.  There is, however, an important sub-population
of highly active RS CVn stars \citep{hall76} that are intrinsically luminous.

Magnetic activity is
governed by the Rossby number \citep{noyes84},
$R_O \equiv P/t_c$, where $P$ is the
rotation period and $t_c$ is the convective overturn timescale.
Greater rotation (smaller $P$) induces faster buildup of magnetic
fields.  Deeper convection (bigger $t_c$) permits the fields to build
up for a longer time before they propagate to the surface.  In the
$R_O$ regime of interest here, the observed rms photometric variability
$A_r$ is a very steep function of $1/R_O$ \citep{hartman09}
\begin{equation}
A_r \propto R_O^{-3.5\pm 0.5};
\qquad
R_O\equiv {P\over t_c}.
\label{eqn:rossby}
\end{equation}

As stars leave the main sequence they will develop deep surface
convection zones as they become cooler, but they will also be
expand substantially and slow down due to angular momentum
conservation.
Hence, some special circumstance
is required to induce or permit relatively rapid rotation.
There are three potential mechanisms.  First, a K dwarf may find
itself in a close binary (either by birth, or through 3-body interactions),
and thus be spun up by tides.  Second, an F or G dwarf may find itself
in a wider binary that is not initially tidally interacting.  But
as the dwarf evolves into a K subgiant, its expanding radius
enables tidal interactions with its companion that then spin up the 
subgiant.  Finally, 
stars in a narrow range of masses, $1.25\,M_\odot\la M \la 1.5\,M_\odot$
(typically F dwarfs) can spend most of their lives spinning fairly
rapidly because of their shallow convection zones and so are still
spinning when they evolve into K subgiants.  The lower mass limit
is required for fast rotation to survive.  Above the upper mass limit,
stars evolve so rapidly through the Hertzsprung gap that they spend
almost no time as subgiants.
The resulting single-star RS CVn subgiants therefore span 
a narrow range of ages,  $7\,{\rm Gyr}\ga t \ga 3\,{\rm Gyr}$.

Here we report the detection of the first spectroscopically
confirmed RS CVn star in the 
Galactic bulge.  The detection was beyond serendipitous.  It resulted from
intensive spectroscopic and photometric observations of an extremely
rare high-magnification microlensing event of a subgiant source.
Only about 1 bulge subgiant per 100 million is so magnified each year.
The intensive photometry was carried out to find planets (orbiting the
lenses), while the high-resolution spectrum of 
MOA-2010-BLG-523S was
obtained to study chemical abundances of bulge dwarfs and subgiants.

MOA-2010-BLG-523S is a subgiant, with a temperature $T\sim 5123\,$K
and surface gravity $\log g=3.6$ 
\citep{bensby11,bensby13}\footnote{The stellar parameters quoted in this
work are taken from \citet{bensby13}.  These values are slightly revised from
the ones originally given by \citet{bensby11}.  Because we discuss
the history (Section \ref{sec:planet})
of how MOA-2010-BLG-523S was recognized to be an RS CVn star, we
report here, for completeness, the \citet{bensby11} parameters that
were available at that time: $T=5250$, $\log g = 4.0$, [Fe/H]$=+0.1$,
$\xi=2.1\,\kms$.}.
As such, either the second or third mechanism of forming RS CVn stars
should apply.  That is,
it is either in a binary that was ``tidally activated'' by the growth
of the primary as it evolved along the subgiant branch, or it is
an isolated, retired F dwarf.  The mere existence of an isolated 
RS CVn star would be evidence for
intermediate-age bulge stars.  Of course, with just one detection,
one could not make a reliable estimate of the fraction of bulge stars
that are of intermediate age.  But there are other lines of evidence
for such a population, including age estimates of microlensed
dwarfs and subgiants \citep{bensby11} and asymptotic giant branch (AGB)
stars \citep{cole02,vanloon03,utten07}.  Thus, it would be of considerable
interest to distinguish between the single-star and binary-star scenarios.
Unfortunately, we find that both scenarios are plausible, given the
available evidence, and so no definitive statement can be made regarding
a putative intermediate-age population.

A major focus of the present paper is the secure identification
of the microlensed subgiant as an RS CVn star.  However, the process
of this discovery is of independent scientific interest.  The event
became a focus of attention because of deviations from standard microlensing
seen over the peak.  The $I$-band lightcurve was quite well fit by
a planetary model, and hence was far ``along the road'' to being published
as a microlensing planet, in which case it would have been only the
14th such planet.  It was really only very small discrepancies that led
to the gradual unraveling of this picture and the recognition that
the deviations at peak are most likely due to magnified star spots rather
than a planet orbiting the lens star.  The fact that irregular variability
due to spots can be fit by planetary microlensing is sobering. 
As we discuss in Section~\ref{sec:planet}, it implies
that great care is required to securely identify microlensing planets
in high-magnification events for cases of low-amplitude signals that
lack clear microlensing signatures.

\section{{Observational Data}
\label{sec:data}}

Microlensing event MOA-2010-BLG-523 (RA,Dec) = (17:57:08.9, $-29$:44:58)
$(l,b)= (0.59,-2.58)$ was alerted by the Microlensing Observations in
Astrophysics (MOA) \citep{bond01,sumi11} collaboration at 
UT 08:46, 21 Aug 2010, and again 26.5 hours later as a potential 
high-mag event that would peak at $A_{\rm max}\sim 70$
in four hours.  In fact the event continued to rise for another
36 hours, which triggered much more intensive observations.  
At UT 16:51 23 Aug, the Microlensing Follow Up Network ($\mu$FUN)
issued a high-mag alert, predicting a peak at UT 02:00 to 04:00 and
on this basis contacted the VLT bulge-dwarf spectroscopy group, advocating
observations in that time interval.  At the same time $\mu$FUN
organized its own continuous photometric observations using the
1.3m SMARTS telescope at CTIO to begin shortly after twilight.
Very importantly in the present context, these
observations were carried out with the ANDICAM camera, which is
equipped with an optical/infrared dichroic, so that it can take
images simultaneously in, e.g., $I$-band and $H$-band.

While the prediction of peak time turned out to be correct,
VLT was unable to observe the event exactly when requested due to
a conflict with technical activities, but did make a 2 hour exposure
(split in 4x30-min) with UVES on VLT beginning near twilight (UT 23:56).  
The main information on this spectrum has already been reported
by \citet{bensby11}.

There are two other very important data sets coming from the
Optical Gravitational Lens Experiment (OGLE) \citep{ews1,ews2}.
The event itself was
monitored by OGLE-IV, which began operations in March 2010.  However,
during 2010, OGLE-IV was in commissioning phase and so did not issue
alerts.  The data were first reduced in November 2010.  Unfortunately,
the target falls in a gap between chips in the new 32-chip OGLE-IV camera,
meaning that the target was captured only when small pointing errors
moved the target onto a chip, which occurred about 1/3 of the time.
Due to the high quality of OGLE data (and despite the reduced coverage),
it was already evident that the source was a low-amplitude
variable and indeed it was checked (and confirmed) at the time of
the image reductions that these variations were not due to chip-edge effects. 
Hence, AU had already suggested at this time that
``the analysis of this object may be more complicated than expected''.

The target also appears in OGLE-III, which took microlensing data from 
2002-2009.  And in addition, it was in a field that was the subject
of a special high-cadence 46 day campaign in 2001 whose aim was
to find transiting planets, during which it was observed 786 times.

In addition there were several other data sets, which in particular
define the falling wing of the lightcurve extremely well.  These include
the RoboNet 2.0m Faulkes North Telescope (SDSS-$i$) in Hawaii, the PLANET 1.0m
Canopus Telescope ($I$) in Tasmania (Australia), the PLANET 0.6m
telescope ($I$)in Perth, Australia, and the following $\mu$FUN
telescopes: Auckland 0.4m ($I$), Farm Cove 0.36m (unfiltered),
Kumeu 0.36m ($I$), Molehill 0.3m (unfiltered) (all in New Zealand),
The 0.6m University of Canterbury B\&C telescope intensively observed
both wings of the light curve. Like the MOA 1.8m telescope, it is located
at Mt.\ John, New Zealand.
Finally the MiNDSTEp 1.5m telescope $(I)$ in La Silla,
Chile obtained data including a few points over the crucial peak region.

\section{{Microlensing Analysis (Simple Version)}
\label{sec:ulens}}

As discussed in Section~\ref{sec:planet}, a complete analysis of
the microlensing event MOA-2010-BLG-523 is complicated by spots
on the surface of the source (called MOA-2010-BLG-523S).  However,
it is possible to derive reasonably robust estimates of all the microlensing
parameters required to constrain the source properties without
detailed modeling of these complexities.

We begin by simply excising the data within 0.8 days of the peak
and fitting the rest of the lightcurve flux $F$ to the standard 
Einstein-Liebes-Refsdal-Paczy\'nski (1936, 1964, 1964, 1986) 5-parameter form
\begin{equation}
F(t) = f_s A(u[t]) + f_{\rm b};
\quad
u^2 = {(t-t_0)^2 + u_0^2\over t_\e^2};
\quad
A = {u^2 + 2\over u\sqrt{u^2+4}}
\label{eqn:pac}
\end{equation}
Here $A$ is the magnification,  $u$ is the projected source-lens
separation in units of the Einstein radius, $u_0$ is the impact parameter,
$t_0$ is the time of closest approach, $t_\e$ is the Einstein crossing
time, $f_{\rm s}$ is the source flux, and $f_{\rm b}$ is any blended flux that
does not participate in the event but is within the same point spread
function (PSF) as the source.  If there is more than one observatory,
then each requires its own $(f_{\rm s},f_{\rm b})$.  We find
\begin{equation}
t_0= 5432.603\pm 0.002;
\quad
t_E=18.5\pm 0.5\,{\rm day}
\quad, 
u_0\la 0.002
\label{eqn:simpleparms}
\end{equation}
and for the OGLE observatory
\begin{equation}
I_{\rm s} = 19.33 \pm 0.03;
\quad
{f_{\rm b}\over f_{\rm s}} =  0.03\pm 0.03.
\label{eqn:fsfb}
\end{equation}
(All times are given in HJD$'=$HJD-2450000).

Inspection of the relatively flat-peaked lightcurve shows that the 
lens crossed directly over the source and that the source
crossing time is (crudely) of order $t_*\sim 0.15\,$day, implying a source 
size (normalized to the Einstein radius) $\rho \equiv t_*/t_\e\sim 0.008$.
Hence, because $u_0\ll \rho$, (and noting that $A\rightarrow u^{-1}$
for $u\ll 1$) we can approximate the peak predicted
magnification as
\begin{equation}
A_\max = \langle r^{-1}\rangle \rightarrow {2\over\rho}
\biggl[1 + \biggl({3\pi\over 8} - 1\biggr)\Gamma\biggr]
\label{eqn:amax}
\end{equation}
where $\langle r^n\rangle$ is the $n$th moment of the source surface
brightness, and where we have assumed a linearly limb-darkened (and
unspotted) source in making the evaluation, in which case the moments
can generally be evaluated
\begin{equation}
\langle r^n\rangle = {\rho^n\over n/2 + 1}(1 - \alpha_n\Gamma);
\quad
\alpha_n = 1 - {(3/2)!(1 + n/2)!\over (3/2 + n/2)!}.
\label{eqn:moments}
\end{equation}
Here $\Gamma$ is the ``natural'' form of the linear limb-darkening
coefficient, defined by surface brightness 
$S(r)\propto 1- \Gamma[1 - (3/2)(1 - (r/\rho)^2)^{1/2}]$
\citep{albrow99}.  It is related
to the standard form $u$ by $\Gamma = 2u/(3-u)$.  It is more ``natural''
in the sense that there is no net flux associated with the limb-darkening
term, which results in simpler formulae when written in terms of $\Gamma$.
This includes not just the moment equations (\ref{eqn:moments}), but
all formulae without exception.  For example, the limb-darkening
term in the standard formula for ellipsoidal variation \citep{morris85},
$(15 + u)/(3-u)$ becomes simply $(5+3\Gamma)$.


We adopt $\Gamma_I = 0.477$ from \citet{claret00}, by applying the
stellar parameters measured by \citet{bensby11}: $T=5123\,$K,
[Fe/H]=+0.06, $\log g = 3.6$, $\xi = 1.68\,\kms$.  Hence, 
$A_\max = 2.17/\rho$.  We evaluate $A_\max$ by taking the ratio
of observed flux at peak to the fit value of $f_{\rm s}$, and get very
nearly the same answer, whether using the average of the two OGLE
peak points, or a median estimate of CTIO near-peak points:
$A_\max = 265$.   We thereby derive
\begin{equation}
\rho = {2.17/A_\max} = 0.0082 \pm 0.0003
\label{eqn:rho}
\end{equation}
where the error is derived from the 3\% error in $f_{\rm s}$ and a 3\% error
in the peak flux due to spots.

\section{{Observational Properties of MOA-2010-BLG-523S}
\label{sec:evidence}}

\subsection{{Baseline Variability}
\label{sec:baselinevar}}

\begin{figure}
\plotone{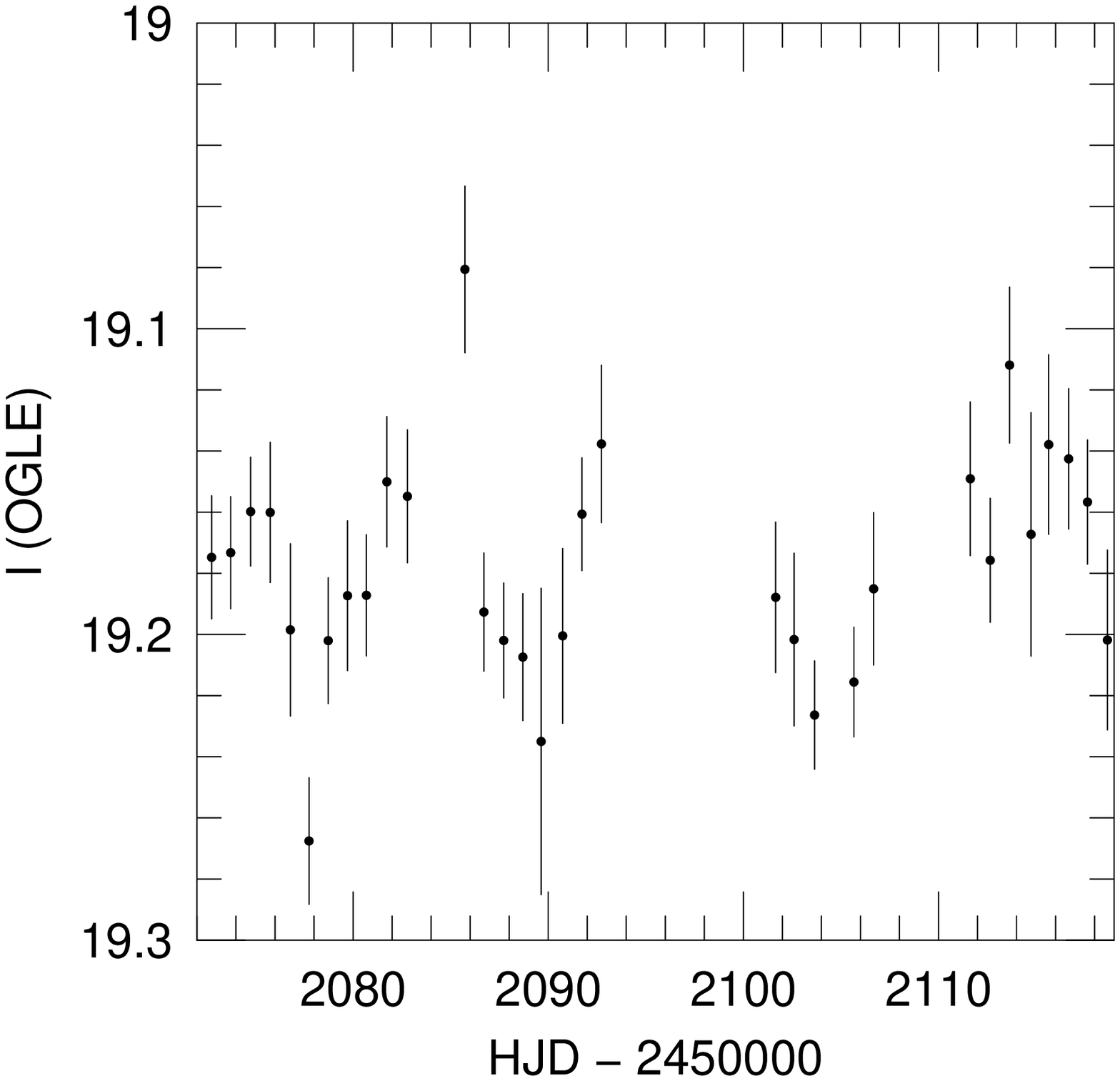}
\caption{\label{fig:2001var}
Lightcurve of MOA-2010-BLG-523S from the 2001 high-cadence OGLE transit
campaign, binned by day.  There are a total of 786 observations on
32 nights, spread over a 46 day interval.  The underlying data have
typical errors of 0.10 mag unbinned, hence 0.02 mag when binned.
The source shows periodic or quasi-periodic oscillations with a period
of roughly 12 days.
}
\end{figure}

As we will argue below, the rms variability of the source is about 3\%.
This is to be compared with the photometric errors, which are typically
close to 10\%.  If the source were a strictly periodic variable, then
the period could easily be identified by folding the light curve,
despite the low signal-to-noise ratio (S/N) of individual points.
The situation is more complex for a quasi-periodic variable (as would
be expected for a rotating spotted star).   We are therefore quite
fortunate that the source lay in a 2001 OGLE transit-campaign field,
which was observed 786 times on 32 separate nights during a 46-day
window.  Binning the data by day, we therefore achieve errors of 0.02,
which is comparable to the amplitude of the signal.  The result is
shown in Figure~\ref{fig:2001var}.  The lightcurve gives the clear
impression of variability with a period of order 12 days.

\begin{figure}
\plotone{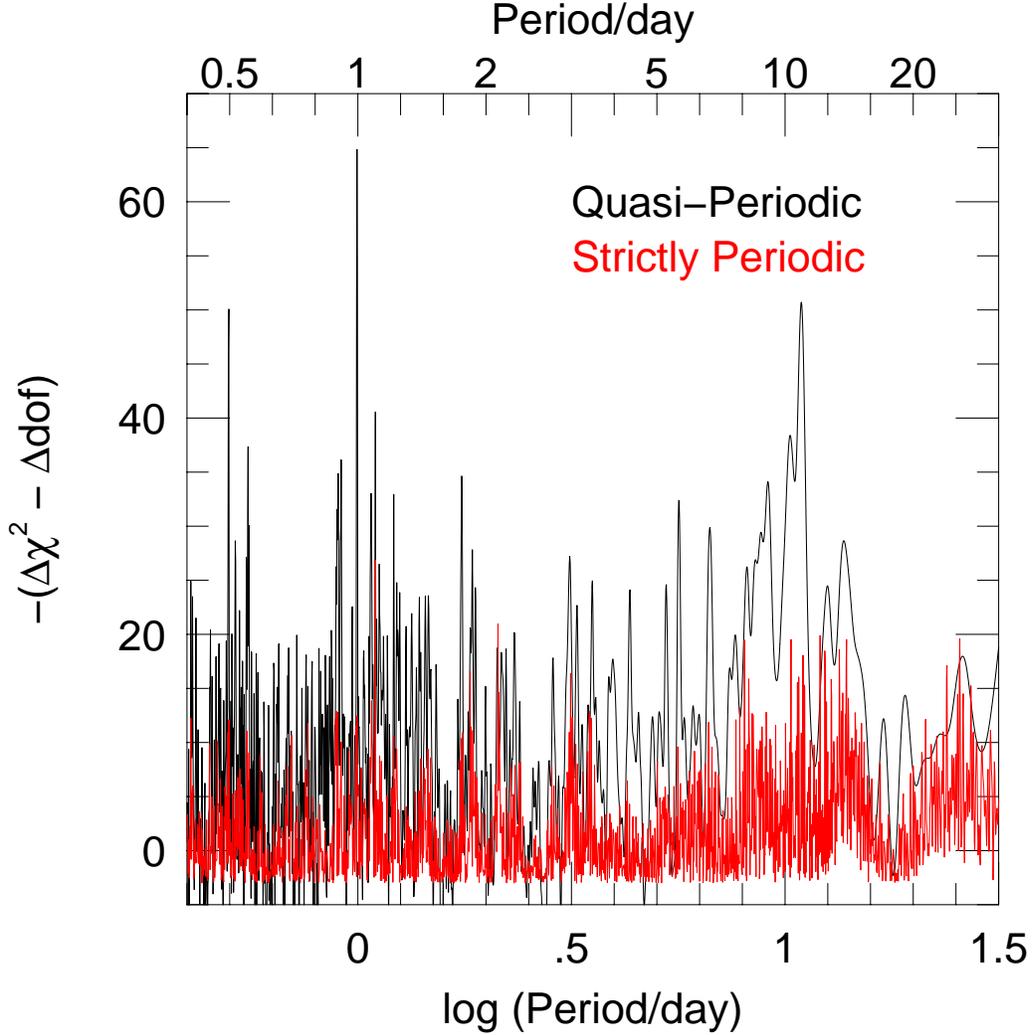}
\caption{\label{fig:logp}
Goodness of fit of strictly periodic (red) and quasi-periodic (black)
models of variability of MOA-2010-BLG-523S over nine OGLE-III seasons
(2001-2009).  The strictly periodic models have 12 free parameters
(period, amplitude, phase, plus zero-point offsets for each season),
while the quasi-periodic models have 28 (additional phases and amplitudes
for each season).  The ordinate shows the difference in $\chi^2$ relative
to a model with 9 parameters (zero-point offset at each season), 
[$\chi^2=2686.75$ for 2531 dof], taking account of the different number
of dof.  Except for a spike very close to 1 day
($0.9947\pm 0.0005$ day) the highest
peak is at $P_0=10.914\pm 0.055$ day.  The quasi-periodic models are
clearly favored over the strictly-periodic ones.  Other notable
peaks are at the alias of the sampling frequency (0.5 day), and
at the aliases of the main peak, $P_\pm = 1/(1/P_0\pm 1/{\rm Day}_{\rm synod})$
= (0.913,1.098) day.
}
\end{figure}

We then use all the OGLE-III data to test for a quasi-periodic
signal.  If this is a spotted star, we expect that the underlying
physical mechanism (rotation of the star) will be strictly periodic,
but that the phase of the variations will drift over time as
spots appear and disappear.  As discussed above, except during the
transit campaign, we are compelled to fold the data to pick up
any signal at all.  On the other hand, if we fold data over an
interval that is too long, the result will suffer from destructive
interference between different phase regimes.  We therefore
consider separate fits to the data for each of the nine seasons,
2001-2009.  In each trial, we hold the period fixed at a common
value for all seasons.  Hence, there are 28 parameters
[Period + $9\times$(phase, amplitude, zero-point)].  At 
$P=10.914\pm 0.055$ days, there is an improvement of 
$\Delta\chi^2=69$ relative to a fit for constant magnitude
in each season (9 parameters), i.e., 19 fewer parameters.  
See Figure~\ref{fig:logp}.

We find
that the phases are not consistent from one season to the next,
suggesting that the variations are not strictly periodic.  To
further test this, we fit for a single phase and amplitude together with a
zero-point offset for each season,  This produces an improvement
(relative to no periodic variations) of only 30 for 3 dof.
Clearly the quasi-periodic variations are favored over strictly-periodic
variations.

\subsection{{Source is the Variable}
\label{sec:sourcevar}}

Faint sources in crowded fields are usually blends of several stars rather than
discrete sources.  And, of course, for microlensing events there
is guaranteed to be at least one additional star along the line
of sight in addition to the source, namely the lens.  Hence, observing
baseline variations does not in itself prove that the source is
variable.  However, from the microlens fit presented in 
Section~\ref{sec:ulens}, we know that the blend is at least 15
times fainter than the source.  Thus, if it were responsible for
the $\sim 3\%$ variations seen at baseline, it would itself
have to vary at the $\ga 50\%$ level on $\sim 11$ day timescales.
Such stars are extremely rare.  Moreover, the chance is remote that one of these
would happen to align with a source that (from other evidence we
will present below) is expected to be variable.  Therefore
we conclude that it is MOA-2010-BLG-523S that is varying.

\subsection{{Calcium H\&K Emission}
\label{sec:calcium}}

\begin{figure}
\plotone{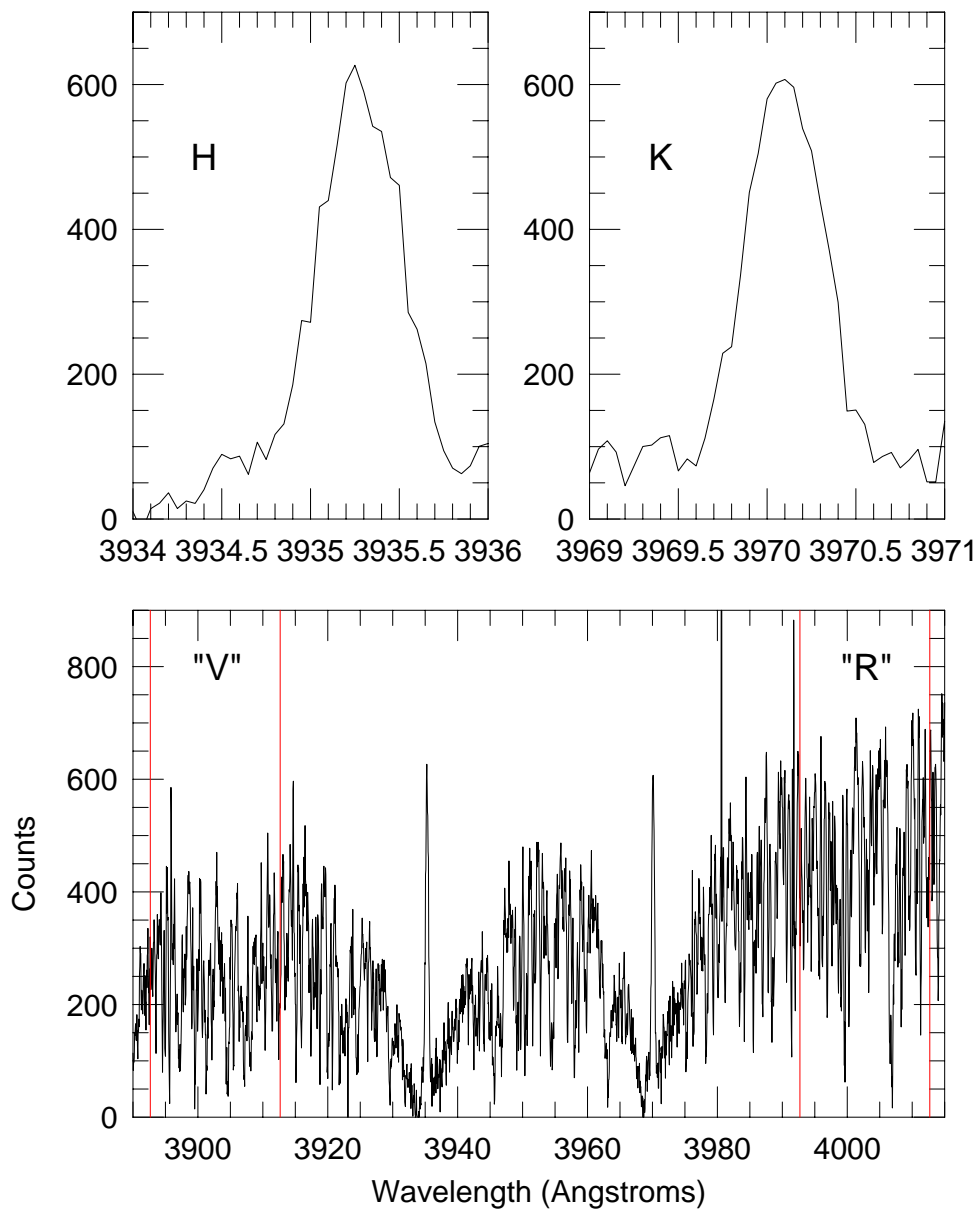}
\caption{\label{fig:blue}
Lower panel: \citet{bensby11} UVES spectrum of MOA-2010-BLG-523S 
in the region of the calcium H \& K lines.  The mean counts 
per $0.05\,$\AA\ in V and R ``continuum'' passbands are 
246 and 439, respectively.  Upper Panels: Zooms of the cores of
the calcium H \& K
lines.  These have total counts of 5596 and 5297 respectively.
Hence the $S$ parameter is $S= (5596+5297)/[(246+439)/0.05\,$\AA]
or $S=0.79\,$\AA.
}
\end{figure}

Figure~\ref{fig:blue} shows the region of the calcium H\&K lines 
in the UVES spectrum taken by \citet{bensby11} 
near the peak of the event.  The emission is extremely strong.
We measure $S_{\rm HK}=0.79\,$\AA\ by taking the ratio of the flux
in these lines to the mean ``continuum'' in the neighboring
``V'' and ``R'' regions (see Fig.~\ref{fig:blue}).
For comparison \citet{isaacson10} found only 3 cases of comparable
or larger $S_{\rm HK}$ among 234 ``subgiants'' in their survey of field stars.
See their Figures 11 and 12.  We will discuss these in 
Section~\ref{sec:isaacson} below, but for the moment note that the
\citet{isaacson10} stars are
substantially redder and more luminous than MOA-2010-BLG-523S.

\subsection{{Microturbulence Parameter $\xi$}
\label{sec:turbulence}}

\begin{figure}
\plotone{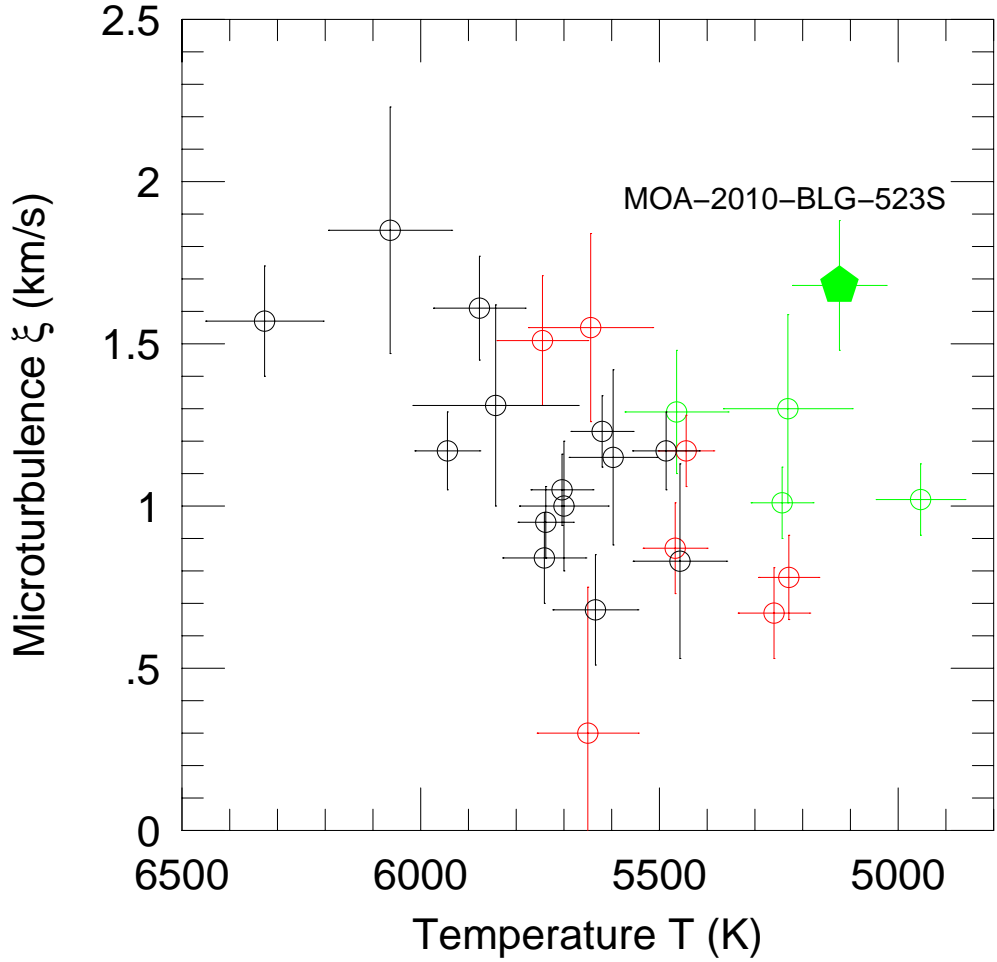}
\caption{\label{fig:xi}
Microturbulence parameter $\xi$ vs.\ temperature $T$ for 26 microlensed
dwarfs and subgiants measured by \citet{bensby10,bensby11}.  
``Dwarfs'' ($\log g>4.2$), ``regular subgiants'' ($4.0\leq \log g <4.2$),
and ``large subgiants'' ($\log g \leq 4.0$) are shown in black, red,
and green respectively.  MOA-2010-BLG-523S is a clear outlier to the
sample.
}
\end{figure}

Figure~\ref{fig:xi} shows the microturbulence parameter $\xi$ plotted
against temperature for 26 microlensed dwarfs and subgiants as
found by \citet{bensby10,bensby11}.  MOA-2010-BLG-523S
has one of the largest $\xi$.  Moreover, it is well above the upper envelope
of points on
the low-temperature part of the diagram.  This high ``microturbulence'' may
reflect real turbulent motions on the surface of the star (as
would be expected for an active star), but may in part reflect
rotational motion.  Since microturbulence represents a Gaussian velocity
distribution that adds to line is quadrature with other effects,
like instrumental resolution, unmodeled rotational motion will contribute
to $\xi$ as
\begin{equation}
\Delta\xi^2 = {\langle r^2\rangle\over \langle r^0\rangle}
{(v\sin i)^2\over 2}
= {1 - 0.2\Gamma\over 4}(v\sin i)^2
\quad (u\gg\rho)
\label{eqn:xi}
\end{equation}
where $v\sin i$ is the projected rotational motion.

In fact, Equation~(\ref{eqn:xi}) applies to sources that are not
differentially magnified, which is of course the usual case, but not
the present one.  If the lens were directly aligned with the source,
then 
\begin{equation}
\Delta\xi^2 = {\langle r^1\rangle\over \langle r^{-1}\rangle}
{(v\sin i)^2\over 2} \simeq {1 - 0.3\Gamma\over 6}(v\sin i)^2
\quad (u=0).
\label{eqn:xi2}
\end{equation}
i.e., roughly 2/3
of the non-differentially magnified case.  For the actual geometry
at the time of VLT spectra and $I$-band limb-darkening, we find
below that $\Delta\xi^2=0.2(v\sin i)^2$.  Hence the measured $\xi$
places an upper limit on $v\sin i$,
\begin{equation}
v\sin i \la  \sqrt{5}\xi = 3.8\,\kms
\label{eqn:vsini}
\end{equation}

\subsection{{Lithium}
\label{sec:lithium}}

In principle, it is possible to produce an isolated rapidly spinning subgiant
(hence, an isolated RS CVn star) in an old population via stellar mergers.  
For example, a 10-Gyr
solar mass star could begin evolving off the main sequence and 
swallow a smaller star, say $0.3\,M_\odot$, that had been its companion.  
This would both spin up the cannibal and provide fresh fuel to extend its 
life.  The mass would be raised above the break in the \citet{kraft70} curve,
so that the star would not substantially spin down during its
extended life.  It would then evolve along the subgiant branch
in a manner similar to any other $1.3\,M_\odot$ star.  However, this
scenario is ruled out in the present case because \citet{bensby11}
detected lithium with abundance $\log\epsilon({\rm Li})= 1.6$.
Essentially all lithium would have been destroyed if there had been a stellar
collision \citep{hobbs91,andronov06}.  
Thus, if MOA-2010-BLG-523S could be shown to lack companions,
it would be of intermediate age.

\subsection{{Radial Velocity}
\label{sec:RV}}

The fraction of microlensing events toward the bulge whose source
stars lie in the bulge (as opposed to the foreground disk) is $\ga 95\%$.
This is primarily because the optical depth to lensing is much higher
due to the higher column of lenses.  But this effect is also compounded
by the fact that there are simply more bulge sources in these fields
compared to disk stars.  Nevertheless, if a source is weird in some way,
its weirdness may be intrinsically connected with it being one of the
small fraction of disk sources.  This possibility is especially relevant
in the present case because the disk is known to harbor a population
of youngish subgiants, whereas the bulge is not.

The source radial velocity (RV), $v_r = +97.3\kms$ \citep{bensby11},
makes it highly unlikely that it is in the disk 
because the expected value for disk stars is
$v_{r,\rm disk} = +10\pm 34\,\kms$ (compared to $+10\pm 100\,\kms$ 
for the bulge).

\subsection{{Source Size}
\label{sec:size}}

\citet{bensby11} derive an equivalent $(V-I)_0=0.86$ color from
their spectroscopic solution (primarily from the temperature, but also
taking account of
the metallicity and gravity).  We find from the microlens solution
in Section~\ref{sec:ulens} that the unmagnified source flux is
$\Delta I=3.18$ mag fainter than the clump.  From the color-magnitude
diagram of the neighboring field, there appears to be very little
differential reddening.  Hence $\Delta I\simeq \Delta I_0$.  Based on
the measured metallicity distribution of bulge stars, \citet{nataf12}
estimate that the absolute magnitude of the
clump is $M_{I,\rm cl}=-0.12$.  Therefore, the absolute magnitude of the
source is 
\begin{equation}
M_{I,\rm s} = M_{I,\rm cl} + \Delta I - 5\log{D_{\rm s}\over D_{\rm cl}}
= 3.06 - 5\log{D_{\rm s}\over D_{\rm cl}}
\label{eqn:misource}
\end{equation}
where the last term is the ratio of the distances to the source and 
the clump.
We then apply standard techniques \citep{yoo04} to evaluate the source
radius, first using \citet{bessell88} to convert $(V-I)\rightarrow(V-K)$
and then using \citet{kervella04} to obtain the $K$-band surface
brightness from the $(V-K)$ color.  Finally, we find
\begin{equation}
R_{\rm s} = 2.15\,R_\odot {D_{\rm s}\over D_{\rm cl}}.
\label{eqn:rsource}
\end{equation}
Note in particular that this derivation is independent of any 
assumption about the Galactocentric distance $R_0$ or the geometry
of the Galactic bar, etc.  Subgiants 
would be expected to have $R_{\rm s}\ga 2\,R_\odot$.  
Hence, the source cannot lie substantially closer than the bulge because
it would then be too small to be a subgiant (as indicated by its
spectroscopic gravity).

\subsection{{Consistency of Spectrum with Stellar Rotation}
\label{sec:rotate}}

If the source is rotating with a period $P=10.9\,$days, as seems
indicated by the quasi-periodic variability seen in Figures
\ref{fig:2001var} and \ref{fig:logp}, then the surface
velocity is $v=2\pi R_{\rm s}/P = 10.0\,\kms (R_{\rm s}/2.15\,R_\odot)$.
The upper limit $v\sin i \la 3.8\,\kms$ (Eq.~[\ref{eqn:vsini}])
then implies $i\la 22^\circ$.  This is a plausible value since
randomly oriented stars will be uniformly distributed in $\cos i$.
That is, 7\% of stars have $i< 22^\circ$, which is small but not
implausibly so.

\subsection{{Consistency with Maoz-Gould Effect}
\label{sec:maozgould}}

\citet{maoz94} predicted that microlensing of rotating stars would
generate an apparent RV shift, which would change during the course
of the event.  Their principal point was that the magnitude of
this effect falls
off only linearly with relative source-lens separation $z\equiv u/\rho$,
compared to the quadratic fall-off of photometric effects:
\begin{equation}
\Delta v = 
{\langle r^2 \rangle\over \langle r^0 \rangle}{v\sin i\over 2 z}\sin\phi
= {1 - 0.2\,\Gamma\over 4z}v\sin i\sin\phi
\qquad (z\gg 1)
\label{eqn:maozgould}
\end{equation}
where $\phi$ is the angle between the source-lens separation and
the projected spin axis.

The \citet{bensby11} spectrum is actually composed of 4 30-minute
exposures, centered on HJD$-2455432=$ (0.510,0.531,0.552,0.573).
Because the lens came very close to source center, we will adopt
$z = (t-t_0)/t_*$.  As we discuss in Section~\ref{sec:planet},
$t_0$ is not very accurately predicted by the lightcurve with the peak
data removed and is actually approximately $t_0=5432.66$
(compared to $t_0 = 5432.60$ found in Section~\ref{sec:ulens}).  Therefore,
at the four epochs, $z=$ (1.00,0.86,0.72,0.58).  We find numerically
that the pre-factor in Equation~(\ref{eqn:maozgould}) at these four epochs
is (0.294,0.333,0.311,0.266).  Thus, the maximum predicted relative
shift is only $0.067 v\sin i\sin\phi < 0.31\,\kms$.   Based on 
cross correlation, the four spectra are consistent at this level.

\section{{Possible Local Analogs}
\label{sec:local}}

We search for local analogs of MOA-2010-BLG-523S in order to better
understand its nature and, to this end, begin by summarizing its
characteristics.

\subsection{{Summary of Characteristics}
\label{sec:chars}}

From \citet{bensby11,bensby13}, 
we know the temperature, iron abundance, gravity,
microturbulence, and lithium abundance: $T=5123\pm 98\,$K, 
[Fe/H]$= 0.06\pm 0.07$
$\log g=3.60\pm 0.23$, $\xi=1.68\pm 0.20$, 
$\epsilon({\rm Li})=1.64\pm 0.10$.  \cite{bensby11} also remark that
the source has high sodium, which they note could be ``fixed'' by making it
500K hotter or increasing $\log g$ by 1 dex.  However, they investigate
these possibilities and reject them.  While \citet{bensby11} did not take
account of differential magnification in their analysis, the impact
of such differential magnification is quite small.  For example, 
``Profile 35'' considered by \citet{johnson10} had much stronger
differential magnification, but this affected the temperature by only 20K
(see their Figs. 3 and 8).

The baseline variability analyzed in Section~\ref{sec:baselinevar}
is best modeled as having constant period $P=10.914\,$days,
but variable phase over
9 years, as would be predicted for a spotted star.  We find
variability amplitudes in these seasons (in mmags) of
$20\pm 4$, $40\pm 8$, $25\pm 10$, $26\pm 9$, $53\pm 9$, 
$28\pm 7$, $13\pm 10$, $27\pm 9$, $83\pm 19$.  Hence,
a median of 0.027 mag, implying an rms variability of 2\%.
This is actually a lower limit, since each season's variability
measure can be impacted by destructive interference between
spot cycles at different phases.

Finally, in Section~\ref{sec:calcium}, we measured calcium H \& K 
emission of $S_{\rm HK} = 0.79\,$\AA.

\subsection{{Comparison by Rossby Number to M37 Sample}
\label{sec:rossby}}

From stellar models, we find that the convective overturn timescale
for a $T=5123$K subgiant is 3 times longer than for the Sun,  while
the measured period is 2.3 times shorter.  Hence the Rossby number
is 7 times higher.  From Figure 17 of \citet{hartman09}, we observe
that stars in M37 with similar $R_O\simeq 0.3$ have rms variability
in the range of 1--6\%.  Hence, the observed variability is quite
consistent with locally observed stars.

\subsection{{Comparison to Isaacson \& Fischer Sample}
\label{sec:isaacson}}

As discussed in Section~\ref{sec:calcium}, \citet{isaacson10} found
only three stars with comparable or greater $S_{\rm HK}$ in their
sample of 234 subgiants.  These are Hipparcos stars HIP 5227, 8281,
and 97501, which have $(V-K)=$ 2.43, 2.56, 2.59.
They are thus considerably redder than MOA-2010-523S, which has
$(V-I)_0= 0.86$ estimated from its spectrum \citep{bensby11}, 
corresponding approximately to $(V-K)_0=1.92$.  They are also 
about 1-1.5 mag more luminous than the subgiants in the \citet{bensby11}
sample shown in Figure~\ref{fig:xi}.  Indeed stars of this color and
luminosity would not be deliberately selected for the \citet{bensby11}
``dwarf and subgiant'' program, and are excluded from the analysis if 
they are observed by accident (Johnson et al., in prep).  Based on their $V/K$
photometry, Hipparcos distances, combined with the \citet{kervella04} surface
brightness relations, these three stars 
have radii of 5.4, 5.9, and 7.0 $R_\odot$, respectively.

All three of these stars are spectroscopic binaries and at least
the first two are broadly
consistent with being tidally synchronized.  Their binary periods
are respectively 27.3 and 30.1, days \citep{eker08}, which would 
imply surface velocities of $10\,\kms$ in both cases, while
\citet{isaacson10} report $v\sin i$ of 14 and $6.2\,\kms$, respectively.
However, D.\ Fischer (2011, private communication) notes that the
profile of the first star is contaminated by lines from a companion,
which she estimates to broaden the $v\sin i$ determination by 20--30\%,
making both stars quite consistent with tidal synchronization.
D.\ Fischer also notes that the third star (Hip 97501) is a clear
double-lined spectroscopic binary, so that its binary nature is not
in doubt even though the 3 RV measurements by \citet{isaacson10} show 
a scatter of only $0.2\,\kms$.

A plausible scenario for these stars is that their moderately
close companions only started to spin them up when they began to
expand their envelopes as they approached the giant branch.
In particular, for the first two, their known periods ($P\sim 30\,$days)
are too long for tidal synchronization while the stars were on the
main sequence.  However, as they evolved along the subgiant branch,
they were clearly tidally spun up, as evidenced both by their
$v\sin i$ and their calcium H \& K activity.

Because MOA-2010-BLG-523S is a much smaller star, a much closer
companion would be required to tidally couple with it.  On the
other hand, its period is much shorter.  Since tidal amplitudes
scale $\propto R^3/P^2$, a companion in an 11 day period could
provide tidal interactions only a factor $\sim 2$ smaller than these
local analogs.  Because MOA-2010-BLG-523S's period is shorter and its phase of
subgiant evolution is longer, it would have many more periods to
tidally synchronize.  Hence, tidal spin-up by a binary companion is a
very plausible explanation for its variability and the strength of its calcium
lines.

\subsection{{Comparison to {\it Kepler} Sample}
\label{sec:kepler}}

\citet{chaplin11} find a dramatic drop in the interval 
5150K$<T<$5400K in the fraction of {\it Kepler}
asteroseismology targets for which they can measure oscillations.
These show variability in the range 0.1--10 mmag, which is modestly
higher than neighboring temperature ranges.  See their Figure~1.
The most plausible interpretation is that subgiants in this
temperature range preferentially acquire spots that physically
interfere with the propagation of stellar oscillations.  (Oscillations
in dwarf stars at these temperature ranges would be undetectable
in any case.)  This cannot be due to close binary companions because
the phenomenon is nearly universal, whereas only a few percent
of stars have such close companions.  Rather, the physical mechanism
must be that as single stars evolve redward on the giant branch,
their convection zones deepen, so $t_c$ increases, and they become
more spotted.  After they pass through the most-affected temperature
range, they expand rapidly, thus increasing their moment of inertia
and so slowing their rotation.  Of course, the more rapidly they
are rotating at the outset, the higher the level of activity, but
the increase in activity at this temperature range is nearly universal.

Because the temperature $T=5123\,$K of MOA-2010-BLG-523S is at the edge
of this affected range, it is also a plausible candidate to be
a non-binary active subgiant.

\section{{Birth and Death of a Microlens ``Planet''}
\label{sec:planet}}

\begin{figure}
\plotone{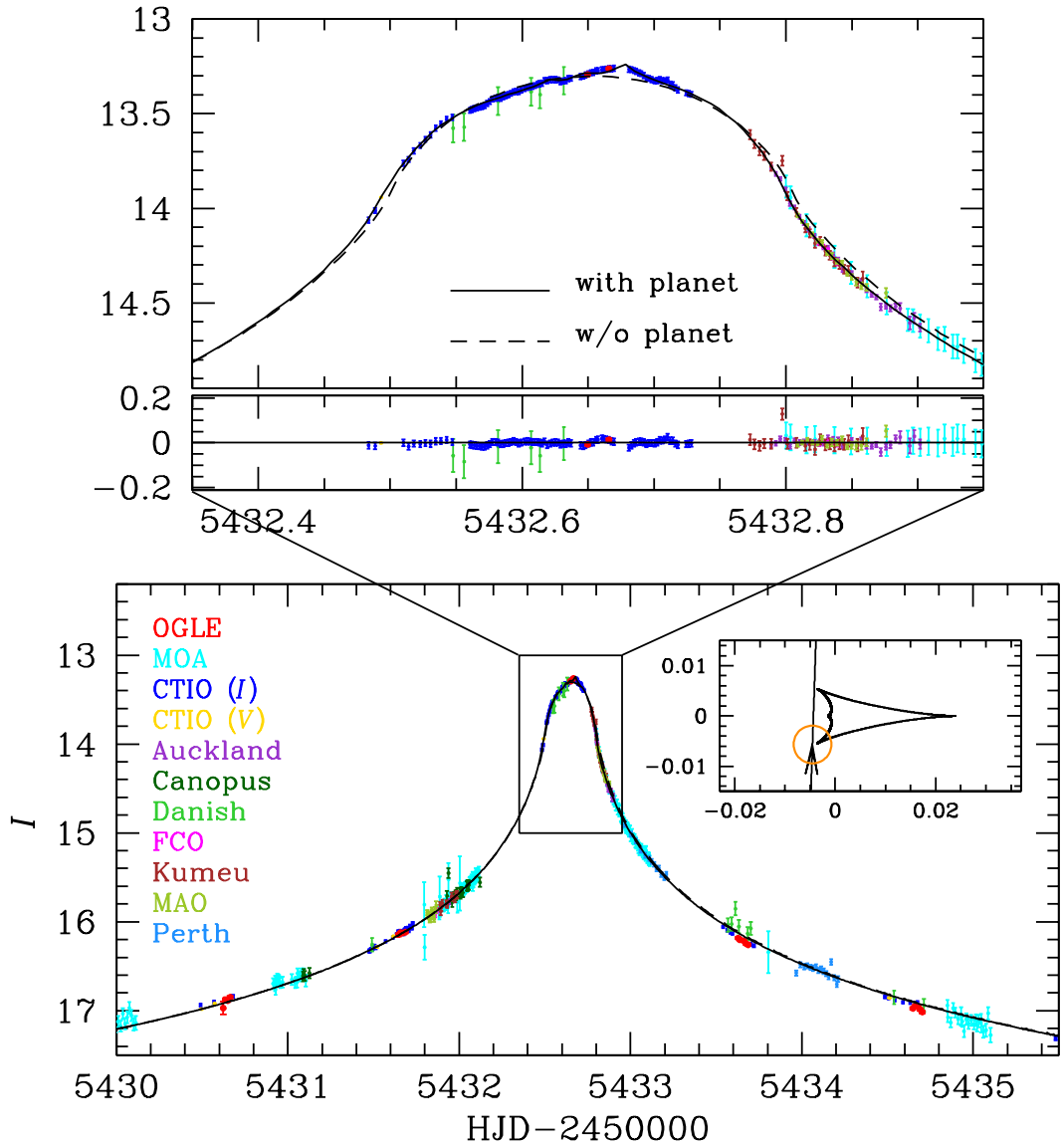}
\caption{\label{fig:planet}
Planetary model of MOA-2010-BLG-523 (black) fit to $I$-band data points
from several observatories as indicated in legend.  $H$-band data are not
shown.
}
\end{figure}

Due to its predicted high-magnification, MOA-2010-BLG-523 was monitored 
almost continuously over peak by the 1.3m SMARTS telescope, although
there were short gaps to check on another, possibly interesting, event.
These data, by themselves, display a significant ``bump'' near peak.
Moreover, the time of the observed peak is asymmetrically
offset from that expected
based on the MOA data (roughly a half day on either side of peak) by
about 1.5 hours. Such bumps and asymmetries are just the type of features
we look for to identify planetary anomalies due to central caustics 
in high-magnification events.  
Within 2 days, preliminary models were circulated and within 4 days,
a planetary model was found that matched all the major lightcurve
features.  See Figure~\ref{fig:planet}.

In accord with standard microlensing practice, one person (JCY) was
assigned to systematically review all the evidence and propose
a final model, which would then be vetted by all groups contributing
data.  Her report stated that the
observed deviations were most likely due to either systematics in the
data or stellar variability and so most likely implied that there
was no planet or, in any case, that it was impossible to reliably
claim a planet.  Note that
none of the evidence presented here that MOA-2010-BLG-523S is
an RS CVn star entered into JCY's reasoning or report.

Rather, JCY was led to question the planetary model because of
three features. First,
the model source crossing time was almost exactly half the naive time
derived from inspection of the lightcurve.  To enable this, the
model has the source pass the planet-star axis almost exactly
at right angles, so that it passes the middle (weak) cusp almost exactly at
peak.  See Figure~\ref{fig:planet}.
Second, one of the three predicted features of the peak
lightcurve takes place in a small gap in the data.  Third, another
feature is somewhat more pronounced in the model than in the data.
Each of these is, by itself, quite plausible and within the
range of microlensing experience, but together were suspicious. 

\begin{figure}
\plotone{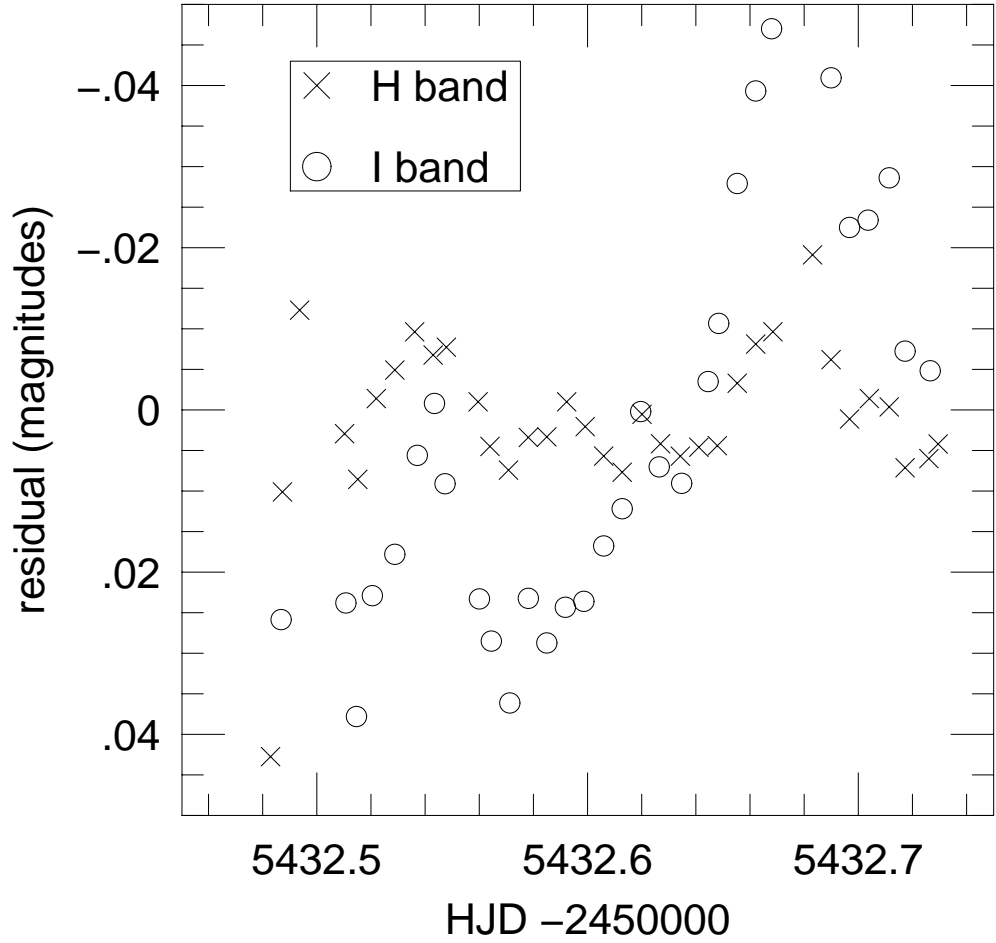}
\caption{\label{fig:res}
Residuals to a point-lens (finite source) fit that uses only $H$-band
data over the peak, for CTIO $H$-band (crosses) and $I$-band (circles).
Points are binned in 10-minute intervals.  Error bars (not shown) are
slightly smaller than points.  In contrast to the $I$-band data, the
$H$-band data show no convincing case for strong deviations.
}
\end{figure}

Hence, JCY sought confirmation of the planetary signal
in other data sets.  Unfortunately, the
two other Chile observatories that might have taken such data 
(OGLE and La Silla) had very sparse coverage.  She therefore
investigated the CTIO ANDICAM $H$-band data, which are normally of
substantially lower quality than the $I$-band data and so are
usually used only for special purposes, such as comparison with
high-resolution post-event $H$-band imaging (e.g., \citealt{mb08310})
or when $I$-band data are saturated (e.g., \citealt{mb07400}).
In this case, the $H$-band data showed a smooth peak, which is
clearly inconsistent with the ``bump'' seen in $I$-band.
See Figure~\ref{fig:res}.  (Note, however, that the $H$-band
peak is still asymmetrically offset by 1.5 hours compared to the time expected
based on data in the wings.)

It is in principle possible to have ``sharper'' features
in $I$-band than $H$-band due to limb-darkening effects.  
This is contrary to the
general expectation that microlensing is achromatic since
in general relativity geodesics do not depend on wavelength.
This exception occurs when
the lens resolves the source because limb darkening is more severe
in bluer passbands making the light profile more compact.
Nevertheless, the amplitude of the difference seen in this
event is much too big to be explained by this effect.  The relative
difference in effective source sizes is
\begin{equation}
\sqrt{\langle[r(H)]^2\rangle\over\langle[r(I)]^2\rangle}-1
=\sqrt{1 - 0.2\Gamma_H\over 1 - 0.2\Gamma_I}-1
\sim 0.1(\Gamma_I-\Gamma_H) <0.02
\label{eqn:hidff}
\end{equation}
whereas the difference in timescales of the observed deviations
is a factor $\sim 2$.
Hence, there is no plausible reason for the difference
in $I$ and $H$ band over the peak.
Moreover, JCY found that the other part of the ``planetary signal'',
the asymmetry in the lightcurve (both $I$-band and
$H$-band), can be fit by ``xallarap'' (orbital motion
of the source about a companion) with periods of 3--15 days.

The path to gathering the evidence summarized in Section~\ref{sec:evidence}
was circuitous.  First, in response to JCY's report, AU reiterated
that the source (or at least some star in the aperture) was a variable
with an 11 day period and a 1\% amplitude.  This variability
had previously been ignored
in the analysis due to the fact that the ``planetary'' deviation
had a much shorter timescale.  Then AG learned from stellar-interiors 
expert MHP that
RS CVn stars were found very frequently at $T\sim 5250\,$K.  MHP then
suggested that the Rossby number scalings from \citet{hartman09} 
could explain the observed variability.  
In the meantime, it was found that variability at fixed
period but random phase (characteristic of spots) is
strongly favored over a strictly periodic signal.  These results
were consistent with an RS CVn star, so one would expect to see
the strong calcium H \& K emission characteristic of such stars
in the UVES spectrum.  However, \citet{bensby11} had not remarked upon
this because the blue spectral
channel is rarely if ever examined for stars in this program because
they are very heavily reddened.  A check of the blue channel indeed 
showed strong H \& K lines.  These lines proved to be easily detectable
in this case not only because they are 
intrinsically strong, but also because the
spectrum was taken at $I\sim 13.3$, which is substantially brighter
than is typical for the \citet{bensby11} sample (see their Fig.\ 1).

In brief, the contradiction between the optical and IR lightcurves
proved to be the crucial turning point in debunking the ``planet'', 
even though a more detailed
investigation of the available data provides overwhelming evidence that
this was a microlensed RS CVn star.

This history argues for caution in the interpretation of planetary
signals, particularly when they are both of small amplitude
and without the discontinuous slopes characteristic of caustic crossings
(e.g., \citealt{ob05169}). 
One may counter in this case that RS CVn stars are extremely rare, 
but the fact remains that this ``rare event'' occurred within
the first dozen or so microlensing planets.  Such rare events in
small samples remind us to be vigilant about our assumptions.

We note that the misinterpretation of microlensed spots as planetary
signals was suggested more than a decade ago by \citet{heyrovsky00},
who specifically cautioned on the difficulty of distinguishing spots
from planets in high-magnification events and even suggested intensive
multi-band photometry as a means to tell the difference. As they remarked,
such multi-band (optical/IR) photometry had already been advocated by 
\citet{gaudi97} as a means to better characterize planetary perturbations.  
This earlier paper (see also \citealt{gould96})
was the motivation to build the optical/IR dichroic
camera ANDICAM \citep{depoy03}, whose $I/H$ observations of MOA-2010-BLG-523
proved crucial in demonstrating that the deviations were due to spots rather
than a planet.  There were several other early investigations of the
interpenetration of spots and binary or planetary microlensing.  
\citet{han00} argued that spots might be easier to study in binary-lens than
single-lens microlensing because the caustics were more likely to transit
the source.  And \citet{rattenbury02} made a broader investigation of
whether spots could in fact be mistaken for planets, arguing that this
was really only possible in the rare events (such as MOA-2010-BLG-523)
in which the lens passes very close to or over the source.  However,
to our knowledge, there are no previously published observations of
microlensed spots.

\section{{RS CVn Stars in the Galactic Bulge}
\label{sec:rscvn_bulge}}

As we have emphasized, it will be quite rare that an RS CVn star
is magnified sufficiently to get a high S/N spectrum of the
heavily extincted Ca H \& K lines.  Nevertheless,
there are other paths toward identifying bulge RS CVn stars.  
\citet{udalski12} found optical counterparts to X-ray sources
from the \citet{jonker11} Galactic Bulge Survey (GBS) catalog,
including 81 spotted stars, which are very probably RS CVn stars.
However, because the underlying X-ray catalog is confined to
$1<|b|<2$, it is likely that a large fraction of these are in the
disk.

There are two relatively straightforward ways to distinguish between
bulge and disk membership.  First, a subset of seven of these 81 stars
are eclipsing.  All but one of these are relatively bright $12.7<I<15.2$,
and so it should be possible to obtain spectra and thus
measure their distances using the method of eclipsing binaries.  Even
the faintest of these, at $I=17.4$, is not beyond reach.  A large
fraction of the remainder could be put on a clump-centric color magnitude
diagram \citep{nataf11}.  In most cases, this should clearly distinguish
between bulge and foreground stars.  Unfortunately, the extinction
map of \citet{nataf12} does not reach most of the \citet{udalski12}
stars because this map is based on OGLE-III data, whereas the GBS survey is
restricted to low-latitude fields that are only covered by OGLE-IV.
However, it should be possible to apply the clump-centric method
to these OGLE-IV fields as well.

\section{{Conclusions}
\label{sec:conclude}}

The evidence presented in Sections~\ref{sec:evidence} 
and \ref{sec:local} that MOA-2011-BLG-523S is an RS CVn star
is overwhelming.
This star shows quasi-periodic variations (the form expected for spots)
with a period of $P=10.9\,$days.  The amplitude of variation (few percent)
is consistent with what one would expect from its inferred Rossby number.
It exhibits very strong calcium H \& K emission, such as is seen
in only 3 out of a sample of 234 local subgiants.  All three are
spectroscopic binaries, and two are known to have periods of 
$P\sim 30\,$days, which given their radii and measured $v\sin i$
implies that they are tidally spun up.  MOA-2010-BLG-523S has 
high microturbulence measured compared to the 
26 microlensed dwarfs and subgiants, particularly among stars
of similar temperature.

Unfortunately, it is not possible to say definitively whether
MOA-2010-BLG-523S is in a binary or not.  Its period relative to
its radius is suggestive of being in the same class of tidally
spun up binaries that includes the 3 calcium-active subgiants just
mentioned.  But its temperature is near the range of
active subgiants found from Kepler seismology, the great majority  
of which must be single stars (or widely separated, non-interacting binaries).
If a strong case could be made that this was not a binary, then
from the lithium measurement (Section~\ref{sec:lithium})
this would be evidence for an intermediate age population.  But
this is not the case.  The fact that the peak is offset from
the time expected from the wings by 1.5 hours strongly suggests
``xallarap'' (orbital motion of the source due to a binary companion). 
However, the irregular character
of the lightcurve, probably due to microlensed spots, compromises
our ability to make a rigorous microlensing fit for xallarap.

The $I$-band lightcurve is well fitted by a planetary-lens model.
The path to discovering that this is a coincidence, and that
the lightcurve anomaly is due to spots was quite circuitous,
as described in Section~\ref{sec:planet}.
This argues for caution in the interpretation of planetary microlensing
events in which the deviations are small and lack
features that are obviously due to a 2-body lens.

\acknowledgments

AG and JCY acknowledge support from NSF AST-1103471. 
Work by JCY was supported by an SNSF Graduate Research Fellowship under
Grant No. 2009068160.  AG, BSG, L-WH, and RWP
acknowledge support from NASA grant NNX12AB99G.
Work by C. Han was supported by Creative Research Initiative
Program (2009-0081561) of National Research Foundation
of Korea.
TB was funded by grant No. 621-2009-3911 from The Swedish Research Council.
Work by S. Dong was performed under contract with
the California Institute of Technology (Caltech) funded by NASA
through the Sagan Fellowship Program.
Work by B.\ Shappee and J.\ van Saders were supported by
National Science Foundation Graduate Research Fellowships.
The MOA  project acknowledges grants 20340052 and 22403003 from JSPS.
T. Sumi acknowledges support from JSPS23340044.
The OGLE project has received funding from the European Research
Council under the European Community's Seventh Framework Programme
(FP7/2007-2013) / ERC grant agreement no.\ 246678 to AU.
MH acknowledges support by the German Research Foundation (DFG). DR 
(boursier FRIA) and J. Surdej acknowledge support from the Communaut/'e 
fran\c{c}aise Belgique Actions de recherche concert\'ees -- Acad/'emie 
universitaire Wallonie-Europe. CS received funding 
from the European Union Seventh Framework Programme (FPT/2007-2013) under 
grant agreement 268421. The Danish 1.54 m telescope is operated 
based on a grant from the Danish Natural Science Foundation (FNU).
KA, DMB, MD, KH, MH, CL, CS, RAS and YT would like to thank the
Qatar Foundation for support from QNRF grant NPRP-09-476-1-078.

\vfil\eject

\end{document}